\documentclass[10pt, conference, compsocconf]{IEEEtran}
\usepackage{amsfonts}
\usepackage{amsthm}
\usepackage{subfigure}
\usepackage{tabu}
\usepackage{multirow}
\usepackage{caption}
\usepackage[utf8]{inputenc}

\usepackage{epsfig}
\usepackage{cite}
\usepackage{array}
\usepackage{tabularx}
\usepackage{color}
\usepackage[cmex10]{amsmath}
\usepackage{mdwmath}
\usepackage{mdwtab}

\usepackage{algorithm}  
\usepackage{algorithmic}

\begin{document}
\title{Diffusion and confusion of chaotic iteration based hash functions}

\author{\IEEEauthorblockN{Zhuosheng Lin\IEEEauthorrefmark{1}, 
Christophe Guyeux\IEEEauthorrefmark{2}, 
Qianxue Wang\IEEEauthorrefmark{1}, and 
Simin Yu\IEEEauthorrefmark{1}}
\IEEEauthorblockA{\IEEEauthorrefmark{1}
College of Automation, Guangdong University of Technology, Guangzhou, China\\
Email: zhuoshenglin@163.com, wangqianxue@gdut.edu.cn, siminyu@163.com}
\IEEEauthorblockA{\IEEEauthorrefmark{2}
Femto-st Institute, University of Bourgogne Franche-Comt\'{e}, Besan\c{c}on, France\\
Email: christophe.guyeux@univ-fcomte.fr}}

\maketitle

\begin{abstract}
To guarantee the integrity and security of data transmitted through the Internet, hash functions are fundamental tools. But recent researches have shown that security flaws exist in the most widely used hash functions. So a new way to improve their security performance is urgently demanded. In this article, we propose new hash functions based on chaotic iterations, which have chaotic properties as defined by Devaney. The corresponding diffusion and confusion analyzes are provided and a comparative study between the proposed hash functions is carried out, to make their use more applicable in any security context.
\end{abstract}


\IEEEpeerreviewmaketitle

\section{Introduction}
Hash functions, as one of the key technologies in information security and cryptographic application domain, are widely used in digital signatures, file integrity checking, authentication, password protection, and so on. At the same time, the analysis of hash functions has recently made some  breakthroughs. Xiaoyu Wang and her team presented new collision search attacks on SHA0 and SHA1\cite{wang2004collisions,wang2005finding,wang2005efficient}. These research results not only shocked people, but also encouraged researchers to construct more secure hash functions.

Chaos, with its high sensitiveness to small changes and initial conditions and long-term unpredictable characteristics, has become an important branch of modern nonlinear science and applications. For instance, a lot of one-way hash functions that are based on chaotic characteristics have been recently proposed\cite{wang2008one,guo2009cryptanalysis}. However, through research, we found that most of these chaotic systems are on real domain. Due to the limited-length when realized in computer or digital devices, this will inevitably lead to finite precision effects and result in dynamical degradation of chaotic systems\cite{li2005dynamical}. Such flaws will make the security performance of hash function declines. 

Chaotic iterations (CIs), defined on integer domains, have been proven to achieve a real chaotic system under the definition of Devaney topological chaos\cite{devaney1989introduction}, which solves degradation of chaotic dynamic properties fundamentally. CIs have been applied to pseudorandom number generation, information hiding, symmetric cryptography, and so on\cite{Guyeux2014Introducing,Guyeux2010Improving}. In this article, we intend to construct a one-way keyed hash function with CIs. Then the diffusion and confusion are analyzed.

The remainder of this article is organized as follows. The basic recalls of CIs and hash function are given in Section \uppercase\expandafter{\romannumeral2}. Our CI-based hash function is proposed and reformulated in Section \uppercase\expandafter{\romannumeral3}. Section \uppercase\expandafter{\romannumeral4} shows its experimental evaluation. This research work ends by a conclusion section in which our article is summarized and intended future work is outlined.

\section{Basic recalls}

This section gives some recalls on topological chaotic iterations and hash functions.

\subsection{Chaotic iterations}

Let us first define some notations that are used in the remainder of this article.  $\mathbb{N}$ is the set of natural (non-negative) numbers. The domain $\mathbb{N}^*= \left\lbrace 1,2,3,\ldots \right\rbrace$ is the set of positive integers and $\mathbb{B} = \left\lbrace 0,1 \right\rbrace$. $\left[\kern-0.15em \left[ 1;N \right]\kern-0.15em \right]= \left\lbrace 1,2,3,\ldots,N \right\rbrace$. A sequence which elements belong in $\left[\kern-0.15em \left[ 1;N \right]\kern-0.15em \right]$ is called a strategy. The set of all strategies is denoted by $\mathit{S}$. $\mathit{S^n}$ denotes the $\mathit{n^{th}}$ term of a sequence $\mathit{S}$, $\mathit{X_i}$ stands for the $\mathit{i^{th}}$ components of a vector $\mathit{X}$.

\newtheorem{myDef1}{Definition}
\begin{myDef1}
Let $f :\mathbb{B}^N\rightarrow \mathbb{B}^N$ be a function and  $\mathit{S}\in \mathbb{S}$ be a strategy. The so-called chaotic iterations are defined by:
	\begin{equation}
	\begin{array}{c}
		x^0 \in \mathbb{B}^N,\\
		\forall n \in \mathbb{N}^*, \forall i \in \left[\kern-0.15em \left[ 1;N \right]\kern-0.15em \right],x^n_i = 
			\begin{cases} 
			x^{n-1}_i, & \mbox{if } S^n \neq i \\
			\left( f\left( x^{n-1}\right) \right)_{S^n}, & \mbox{if } S^n = i 
			\end{cases}
	\end{array}{}
	\end{equation} 
\end{myDef1}

In other words, at $n^{th}$ iteration, only the $S^{n}$-th component of vector $x^n$ is updated.

For a given function $f$, let us define a function $F_f:\left[\kern-0.15em \left[ 1;N \right]\kern-0.15em \right]\times \mathbb{B}^N \rightarrow \mathbb{B}^N $ by:
\begin{equation}
F_f\left( {k,x} \right) = \left( {{x_j} \cdot \delta \left( {k,j} \right) + {{\left( {f\left( x \right)} \right)}_k} \cdot \overline {\delta \left( {k,j} \right)} } \right)_{j=1,2,3,\ldots,N}
\end{equation} 
where $\delta \left( {k,j} \right)=0 \Leftrightarrow k = j$. Consider the phase space: $\mathcal{X}=\left[\kern-0.15em \left[ 1;N \right]\kern-0.15em \right]\times \mathbb{B}^N$, and the map defined on $\mathcal{X}$ by: 
\begin{equation}
G_f\left( S,E\right) = \left( \sigma\left( S\right) , F_f\left( i\left( S\right) ,E\right) \right),
\end{equation} 
where $\sigma$ is the shift function that removes the fist term of the strategy. So the chaotic iterations defined in Equ.1 can be described by the following iterations:
\begin{equation}
	\begin{cases} 
	X^0 \in \mathcal{X}\\
	X^{k+1}=G_f(X^k) 
	\end{cases}
\end{equation} 

For given two points $X=\left( S,E\right) ,Y=\left( \check{S},\check{E}\right) \in \mathcal{X}$, we define the distance between these two points by:
\begin{equation}
	\begin{array}{c}
		d \left( X,Y\right) =d_e \left( E,\check E\right) + d_s \left( S, \check {S} \right), \text{where}\\
		\begin{cases} 
		 d_e \left( E,\check {E} \right) = \sum\limits_{k = 1}^N \delta\left( E_k,\check {E}_k \right)\\ 	
		 d_s \left( S, \check {S} \right) = \tfrac{9}{N} \sum\limits_{k = 1}^{\infty} \tfrac{ \| S^k - \check {S^k}\|}{10^k}
		\end{cases}
	\end{array}
\end{equation} 
in which $\left\lfloor {d\left( X,Y \right) } \right\rfloor = d_e \left( E,\check E\right) $ is the Hamming distance between $E$ and $\check{E^k}$. So $d \left( X,Y\right) - \left\lfloor {d\left( X,Y \right) } \right\rfloor = d_s \left( S, \check {S} \right) $ measures the difference between strategies $S$ and $\check{S}$. More precisely, this floating part is lower than $10^{-k}$ if and only if the first $k$ terms of the two strategies are equal. Moreover, if the $k^{th}$ digit is nonzero, then $S_k \neq \check{S_k}$. 

Considering the distance between $d$ on $\mathcal{X}$, it has already been proven that\cite{guyeux2010topological}:
\begin{itemize}
\item $G_f$ is continuous.
\item Iterations defined in Equ.4 are regular.
\item $G_f$ is topologically transitive.
\item $G_f$ has sensitive dependence on initial conditions.
\end{itemize}

Thus, according to the Devaney's definition\cite{devaney1989introduction,banks1992devaney}, $G_f$ is chaotic.

\subsection{Hash functions}

Let $k\in K$ be a key in a given key space $K$. So a function $h_k\left( \right) $ that maps a key $k$ and a binary bit string $x$ to a string of a fixed length $l$ is a Secure Keyed One-Way Hash Function (SKOWHF)\cite{Bakhtiari1995Keyed}, if it satisfies the following properties:

\begin{itemize}
\item Given $k$ and $x$, it is easy to compute $h\left( k,x\right)$.
\item Without knowledge of $k$, it is hard to compute $h\left( k,x\right)$.
\item For any $x$ or given (possibly many) pairs $x$ and $h\left( k,x\right)$, it is hard to compute $k$.
\item For a given $k$, it is hard to find two values $x$ and $y$ such that $h\left( k,x\right)=h\left( k,y\right)$, but $x \neq y$.
\item Length $l$ has to be larger than 128 bits in order to counter birthday attack.
\item Key space $K$ has to be sufficiently large in order to counter exhaustive key search.
\end{itemize}

\section{CI-based hash functions}
Let us now present our hash function $H_h:K\times\mathbb{B}^* \rightarrow \mathbb{B}^N$ which is based on chaotic iterations recalled before. The key $k=\left\lbrace k_1, k_2, prng\_type\right\rbrace $ is in key space $K= \mathbb{B}^{k1}\times\mathbb{B}^{k2}\times\mathbb{N} $. All the steps are described in the following paragraphs.

The first step of the algorithm is to choose the traditional hash function $h$ that we will use in our own hash function. For our implementations, we have chosen MD5, SHA-256, and SHA-512. And the selective traditional hash function determines the length ($N$) of the output hash value. For MD5, $N=128$, for SHA-256, $N=256$, and for SHA-512, $N=512$. 

Then for the input message $x$, we need to transform the it into a multiple normalized $N$ bits sequence. This pre-treatment is similar to the SHA-1 case. After that, the length of the treated sequence $X$ is $L$. 

In the third step, we use $k_2$ as a seed to generate $L$ bits pseudorandom numbers $m$. In our implementation, the Pseudo-Random Number Generator can be Mersenne Twister (MT), Blum Blum Shub (B.B.S.), XORshift, or Linear Congruential Generator (LCG). This is the $prng\_type$ in key space $K$. The generated pseudorandom numbers are used to construct the strategies. As $2^n = N$, we split its sequence to be  $m=S^0S^1\ldots$, where the length of $S^i$ is $n$. Then the strategy is $S=\left\lbrace S^0S^1\ldots\right\rbrace$, where $S^i$ is transformed to the decimal value.

In the forth step, we first transform the input $k_1$ to binary value which length is $N$. Here we split $X$ into $X=\left\lbrace X^0X^1\ldots\right\rbrace$. Each $X^i$ will be combined with $k_1$ using exclusive-or operation. Then  we combine the result with pseudorandom numbers $m$ using exclusive-or operation too. After that, we use this result as the input of traditional hash function $h$.

Lastly, to construct the digest, chaotic iteration of $G_f$ are realized with the traditional hash function output $h(k_1,X,m)$ and strategies $S$ as defined above. The result of these iterations is a $N$ bits vector. It is translated into hexadecimal numbers to finally obtain the hash value. 

So we define the keyed hash function $H_h:K\times\mathbb{B}^{*}\rightarrow\mathbb{B}^{N}$ by the following procedure
\begin{algorithm}[htb]   
\caption{ The proposed hash function $H_h$}   
\label{hash function}   
\begin{algorithmic}[1] 
\REQUIRE ~~\\ 
The key, $k=\left( k_1,k_2,prng\_type\right)\in K $;\\  
The input message $x\in\mathbb{B}^{*}$;\\   
\ENSURE ~~\\ 
Hash value $H$;  
\STATE Transforming $x$ to sequence $X$ which length is $L$;   
\STATE Use PRNG to generate $m$ which using $k_2$ as a seed and construct strategy $S=\left\lbrace S^0S^1\ldots\right\rbrace$ with $m$;
\STATE Use standard hash function to generate hash value $H=h(k_1,X,m)$;
\FOR{$i=1\ldots$} 
\STATE Chaotic iterations, to generate hash value: $H=G_f(S^i,H)$;
\ENDFOR        
\RETURN $H$; 
\end{algorithmic}  
\end{algorithm}  

Thus $H_h$ is a chaotic iterations based post-treatment on the inputted hash function. If $h$ satisfies the collision resistance property, then it is the case too for $H_h$. Moreover, if $h$ satisfies the second-preimage resistance property, then it is the case too for $H_h$, as proven in\cite{Guyeux2014Introducing}.

\section{Experimental Evaluation}
Before discussing diffusion and confusion, we will give some examples of hash values.
\subsection{Hash Value} 
Let us now consider that the input message is the poem Ulalume (E.A.Poe), which is constituted by 104 lines and 3582 characters. The traditional hash function used here will be the MD5. So $N=128$. To give illustration of the keys properties, we will use this hash function $H_h$ to generate hash values in the following cases: 
\begin{itemize}
    \item[] Case 1. $k_1=50,k_2=50,prng\_type$ is B.B.S..
    \item[] Case 2. $k_1=51,k_2=50,prng\_type$ is B.B.S..
    \item[] Case 3. $k_1=50,k_2=51,prng\_type$ is B.B.S.
    \item[] Case 4. $k_1=50,k_2=50,prng\_type$ is LCG.
    \item[] Case 5. $k_1=50,k_2=50,prng\_type$ is MT.
    \item[] Case 6. $k_1=50,k_2=50,prng\_type$ is XORshift.
\end{itemize}

The corresponding hash values in hexadecimal format are:
\begin{itemize}
    \item[] Case 1. D8ED0DDD1A611C1AEDE0915BE2CA91D3.
    \item[] Case 2. 54B7B1E2C2239CF0FBC327D55CFA7BF2.
    \item[] Case 3. 8453BA95FB088DA84219F1AFCD14E9EE.
    \item[] Case 4. 663F90CB4ECD5E8AF53D2760E01491C8.
    \item[] Case 5. 01C142B339413DEF49E7A65FF43A50DF.
    \item[] Case 6. FD2B8ABC6BE956718669D92367E1680A.
\end{itemize}

From simulation results, we can see that any change in key space $K$ seems to cause a substantial modification in the final hash value, which is coherent with the topological properties of chaos.

For a security hash function, the repartition of its hash values should be uniform. In other words, the algorithm should make full use of cryptogram space to make that the hash values are evenly distributed across the cryptogram space. The parameter we use here is the same as in Case 1. In Figure 1a, the ASCII codes are localized within a small area, whereas in Figure 1b the hexadecimal numbers of the hash values are uniformly distributed in the area of cryptogram space.

\begin{figure}[]
\centering
\subfigure[Plain text sequence (ASCII)] {\includegraphics[height=2in,width=3.5in,angle=0]{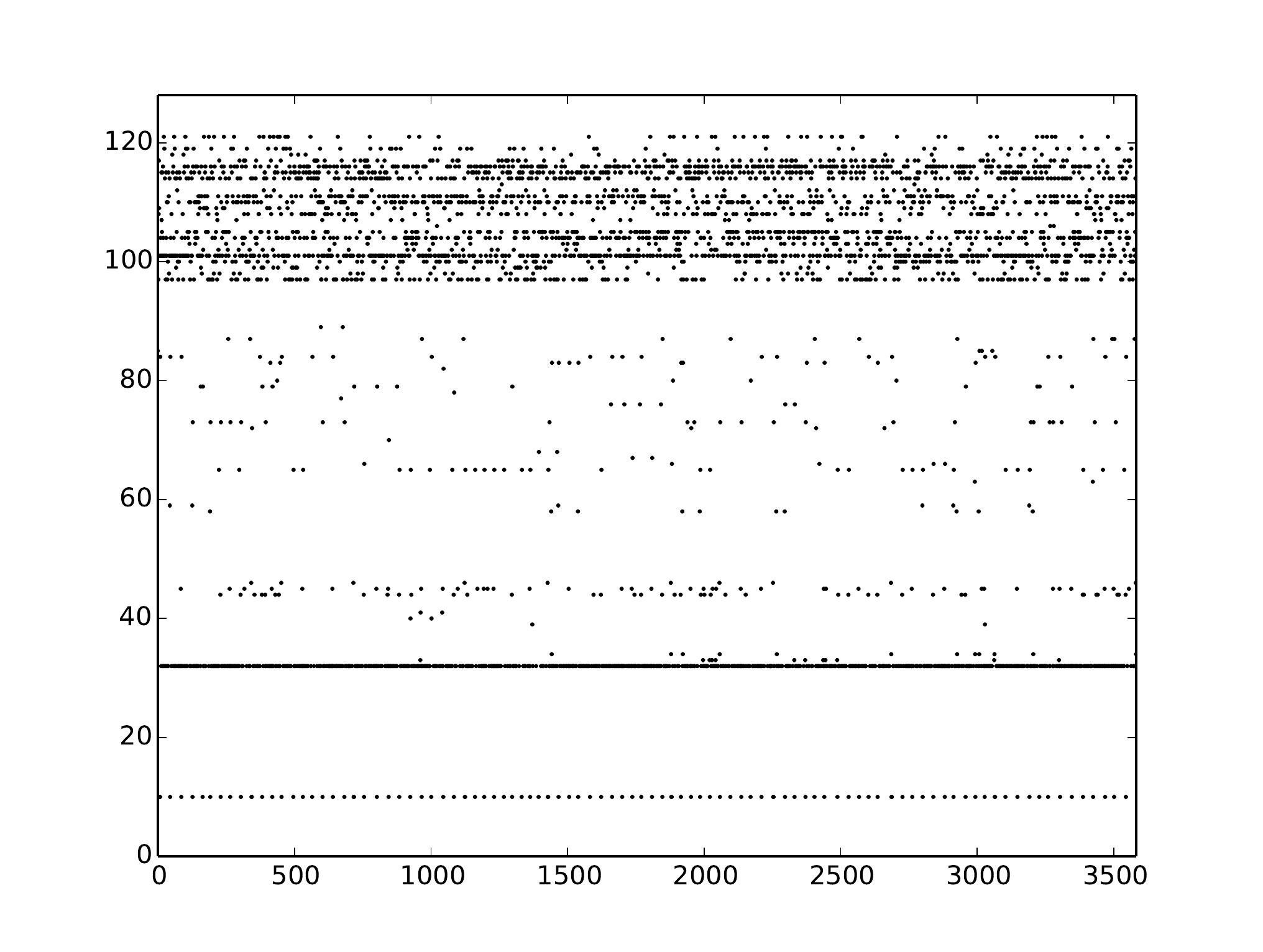}}
\subfigure[Hash value (Hexadecimal)] {\includegraphics[height=2in,width=3.5in,angle=0]{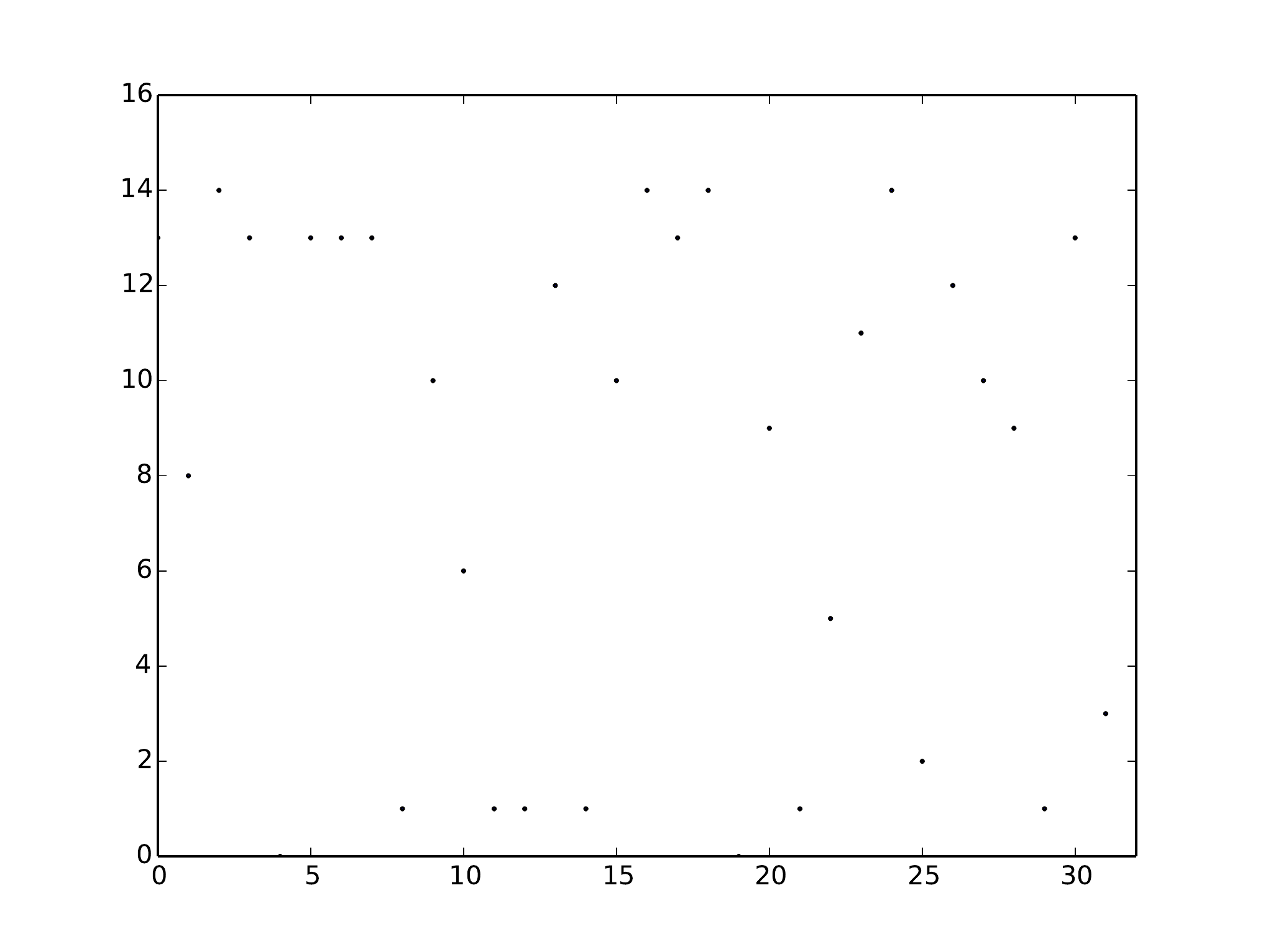}}
\caption{ Distribution of Ulalume poem }
\label{fig1}
\end{figure}

We will now test our hash function with some changes in the input message, and observe the distribution of hash values. The hash function is set with $k_1=50,k_2=50$, and $prng\_type$ is B.B.S.. The hash function used here will be the MD5.
\begin{itemize}
    \item[] Case 1. The input message is the poem Ulalume (E.A.Poe).
    \item[] Case 2. We replace the last point `.' with a coma `,'.
    \item[] Case 3. In ``The skies they were ashen and sober", `The' become `the'. 
    \item[] Case 4. In ``The skies they were ashen and sober", `The' become `Th'. 
    \item[] Case 5. We add a space at the end of the poem.
\end{itemize}

The corresponding hash values in binary format are shown in Figure 2. Through this experiment, we can check that the propose hash function is sensitive to any alteration in the input message, which will cause the modification of the hash value. 
\begin{figure}[]
\centering
\includegraphics[height=2in,width=3.5in,angle=0]{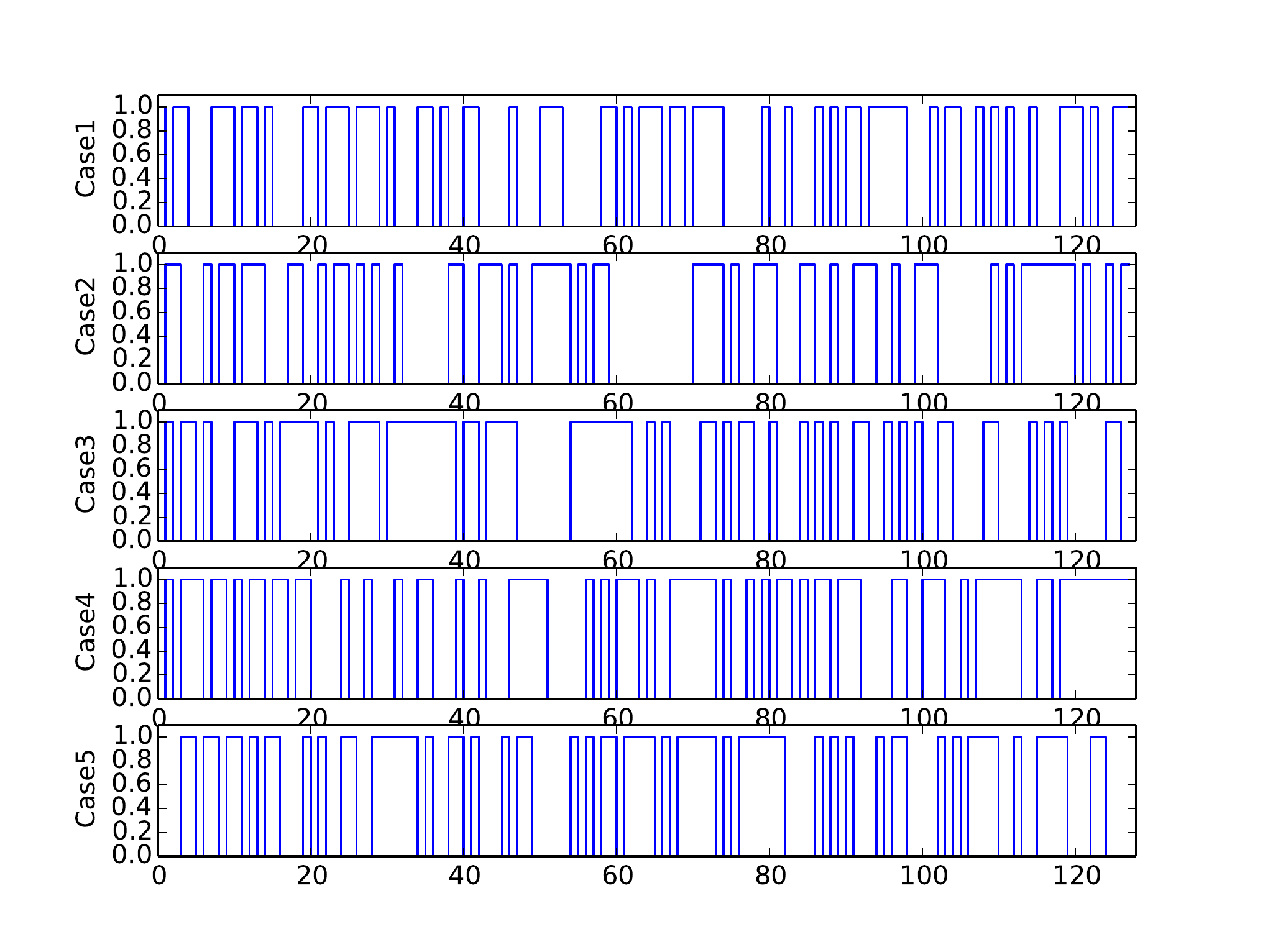}
\caption{ 128 bit hash values in various cases }
\label{fig2}
\end{figure}

\subsection{Diffusion and Confusion}
 In cryptography, diffusion and confusion are two important properties of a secure cipher that has been identified by Claude Shannon in his 1945 classified report ``A Mathematical Theory of Cryptography". Diffusion means that the redundancy of the plain text must be dispersed into the space of cryptogram space so as to hide the statistics of plain text. Confusion refers to the desire to make the statistical relationship between plain text, ciphertext, and keys as complex as possible, which makes attackers difficult to get relation about keys from ciphertext. These concepts are important too in the design of robust hash functions. We now focus on the illustration of diffusion and confusion properties. 
 
 To analyze the statistic of diffusion and confusion, the following common statistics are used:
\begin{itemize}
    \item Mean changed bit number: $\overline{B}=\tfrac{1}{N}\sum_{i=1}^N B_i$.
    \item Mean changed probability: $P=\tfrac{\overline{B}}{L}\times100\%$.
    \item Mean square error of B: $\Delta B=\sqrt{\tfrac{1}{N-1}\sum_{i=1}^N(B_i-\overline{B})}$. 
    \item Mean square error of P:\\ $\Delta P=\sqrt{\tfrac{1}{N-1}\sum_{i=1}^N(\tfrac{B_i}{L}-\overline{P})}\times100\%$, 
\end{itemize}
where $N$ denotes the statistical times, and $B_i$ denotes the changed bits of hash value in $i^{th}$ test, while $L$ denotes the bits of hash value in binary format.

\begin{figure}[]
\centering
\includegraphics[height=2in,width=3.5in,angle=0]{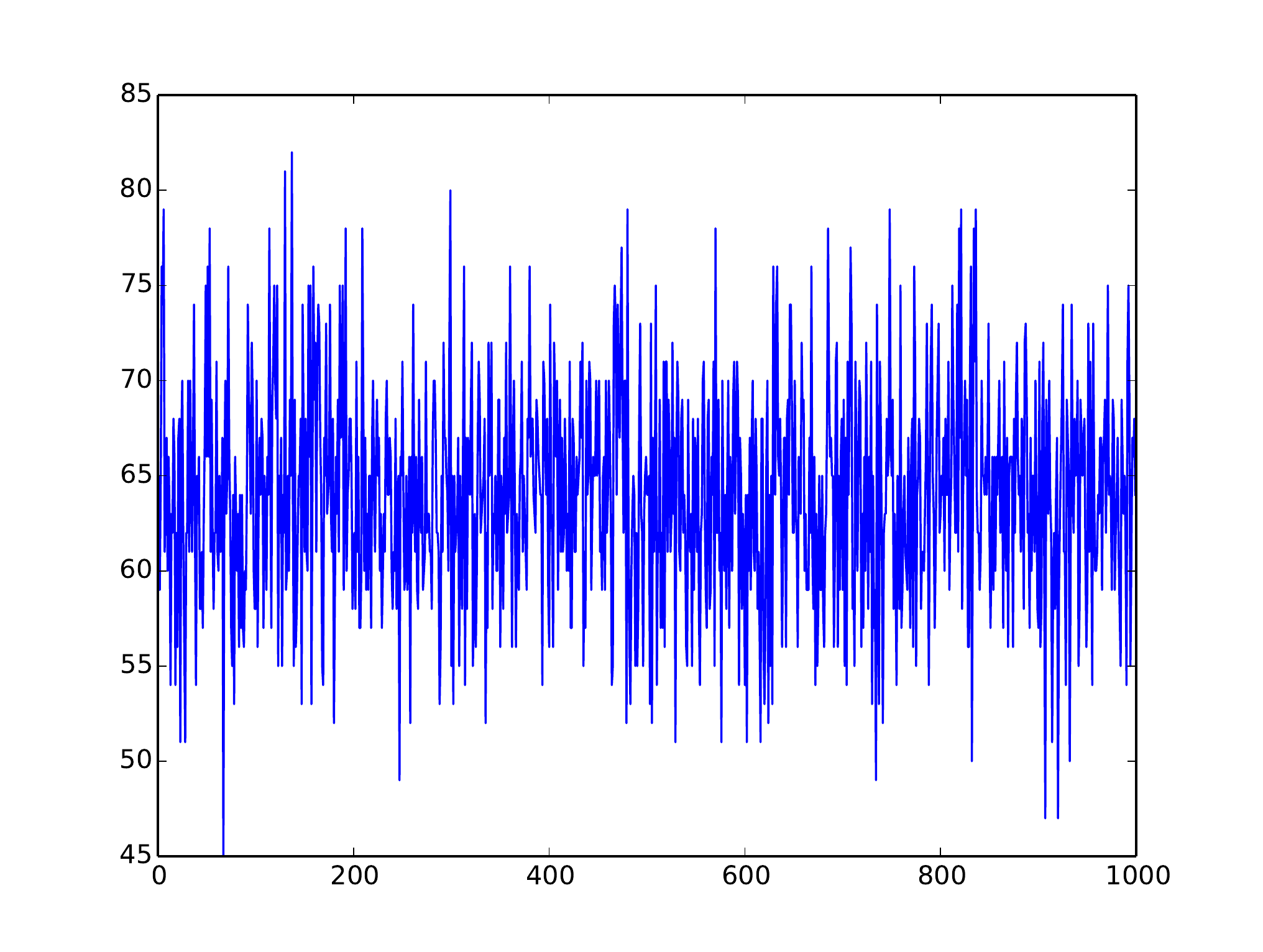}
\caption{Distribution of changed bit numbers $B_i$}
\label{fig3}
\end{figure}

 We use again the poem Ulalume (E.A.Poe) as input message. Using our hash function $H_h$, we will get the original hash value. For this sequence, we toggle only one bit each time. Then we will obtain another hash values. Let $k_1=50,k_2=50$, and $prng\_type$ is B.B.S. The hash function $h$ used is MD5 while test times $N=1000$. The distribution of $B_i$ is shown in Figure 3. From the figure, we can see that a one bit change in the plain text will modify about 64 bits in the 128 bits hash value. In other words, the proposed hash function achieves desired value for such properties.
  
\begin{table}[htb]
\centering
\caption{Statical performance of the proposed hash function}
\label{my-label}
\begin{tabular}{cccccc}
\hline
$prng\_type$                      & $hash\_type$ & $\overline{B}$     & $P(\%)$ & $\Delta B$ & $\Delta P(\%)$ \\ \hline
\multirow{3}{*}{B.B.S}            & MD5          & 64.008  & 50.006  & 5.788      & 4.522          \\
                                  & SHA-256      & 128.085 & 50.033  & 7.880      & 3.078          \\
                                  & SHA-512      & 256.353 & 50.069  & 10.911     & 2.131          \\\hline
\multirow{3}{*}{Mersenne Twister} & MD5          & 63.977  & 49.982  & 5.452      & 4.260          \\
                                  & SHA-256      & 128.316 & 50.123  & 7.858      & 3.070          \\
                                  & SHA-512      & 255.534 & 49.909  & 11.691     & 2.283          \\\hline
\multirow{3}{*}{LCG}              & MD5          & 64.355  & 50.277  & 5.795      & 4.528          \\
                                  & SHA-256      & 128.056 & 50.022  & 7.842      & 3.063          \\
                                  & SHA-512      & 256.106 & 50.021  & 11.539     & 2.254          \\\hline
\multirow{3}{*}{XORshift}         & MD5          & 63.963  & 49.971  & 5.648      & 4.412          \\
                                  & SHA-256      & 127.596 & 49.842  & 8.036      & 3.139          \\
                                  & SHA-512      & 255.955 & 49.991  & 11.573     & 2.260          \\ 
\hline 
\end{tabular}
\end{table}

\begin{table}[htb]
\centering
\caption{Statical performance of the standard hash function}
\label{my-label (stand)}
\begin{tabular}{ccccc}
\hline
standard hash function & $\overline{B}$     & $P(\%)$ & $\Delta B$ & $\Delta P(\%)$ \\ \hline
MD5                    & 63.893  & 49.916  & 5.437      & 4.248          \\ 
SHA-256                & 127.746 & 49.901  & 8.405      & 3.283          \\ 
SHA-512                & 256.084 & 50.016  & 11.232     & 2.194          \\ \hline
\end{tabular}
\end{table}

The desired distribution of hash algorithm should be that small toggle in plain text causes $50\%$ change of hash value. $\Delta B$ and $\Delta P$ show the stability of diffusion and confusion properties. The hash algorithm is more stable if these two values are close to $0$. Observing Table \uppercase\expandafter{\romannumeral1}, both the mean changed bit number $\overline{B}$ and the mean changed probability $P$ are close to the desired value. $\Delta B$ and $\Delta P$ are quite small. Both of them illustrates the diffusion and confusion of our hash function $H_h$ and these capabilities are quite stable. From Table \uppercase\expandafter{\romannumeral1}, we can also know that when $prng\_type = LCG$ or $prng\_type = B.B.S$, all $P$ are larger than $50\%$. But when $prng\_type = LCG$, $\Delta B$ and $\Delta P$ are smaller. To sum up, in our proposed hash function, it is better to choose B.B.S. as pseudorandom number generator. Furthermore, compared with the performance of standard hash functions which is shown in Table \uppercase\expandafter{\romannumeral2}, the proposed one in some situations shows better results.    
 
\section{Conclusion and Future Work}
In this article, a new hash function based on chaotic iterations has been presented. We used pseudorandom number generator to construct a strategy $S$. Then we simulated the proposed hash function's sensitivity to keys and plain text. At last, the performance of diffusion and confusion is discussed. The experimental results show that this hash function is a secure keyed one-way hash function. Through the statical performance of the proposed hash function, we found that B.B.S is a better pseudorandom number generator to construct strategies.

In future work, we will try to apply chaotic iteration to construct pseudorandom number generators. Then we will use this kind of PRNG to construct strategies for hash functions. At the meantime, other properties induced by CIs will be explored.


\bibliographystyle{IEEEtran}
\bibliography{hash_function}
\end{document}